\documentclass[preprint,pre,a4paper,superscriptaddress,nofootinbib]{revtex4}

\usepackage{graphicx}
\usepackage[body={16cm,22cm},bottom=1.5cm]{geometry}
\usepackage{amsmath}
\usepackage{latexsym}
\usepackage{amsfonts}
\usepackage{psfrag}
\usepackage{epstopdf}
\begin{document}
\title{Modelling thermal conductivity and collective effects in a
simple nanofluid}
\author{Mihail Vladkov \footnote{E-mail: mihail.vladkov@lpmcn.univ-lyon1.fr},
Jean-Louis Barrat \footnote{E-mail:
jean-louis.barrat@lpmcn.univ-lyon1.fr}}

\affiliation{
  Laboratoire de Physique de la Mati\`ere Condens\'ee et Nanostructures
               Universit\'e Lyon 1; CNRS; UMR 5586
               Domaine Scientifique de la Doua
               F-69622 Villeurbanne cedex; France}

\date{\today}
\setcounter{page}{1}

\begin{abstract}
Molecular dynamics simulations are used to model  the thermal
properties of a  fluid containing solid nanoparticles (nanofluid).
 The flexibility of molecular
simulation allows us to consider the effects of particle mass,
particle-particle and particle-fluid interaction and that of  the
spatial distribution of the particles on the thermal conductivity.
We show that the heat conductivity of a well dispersed nanofluid
is well described by the classical Maxwell Garnett equation model.
In the case of particle clustering and strong inter particle
interactions the conductivity can be again described by effective
medium calculation taking into account the aspect ratio of the
cluster. Heat transfer is increased when  particles are aligned in
the direction of the  temperature gradient. This kind of
collective effects could be a first step to understand the
substantial increase in the conductivity observed in some
experiments.
\end{abstract}

\maketitle

\newpage
Many experimental studies have suggested that the thermal
conductivity of colloidal suspensions referred to as
``nanofluids'' is unusually high \cite{Eastman,Patel}. Predictions
of effective medium theories are accurate in some cases
\cite{Putnam-poly} but generally fail to account for the large
enhancement in conductivity. In spite of a large number of -
sometimes conflicting or controversial - suggestions and
experimental findings \cite{Keblinski-Cahill},  the microscopic
mechanisms for such an increase remain unclear. One of the
possibilities that was suggested was the effect of Brownian motion
\cite{prasher}, and  appeared to be an attractive and generic
explanation. The essential idea is that the Brownian velocity of
the suspended particle induces a fluctuating hydrodynamic flow
\cite{Alder,Keblinski-flow}, which on average influences
(increases)
 thermal transport. This mechanism is different from transport of
 heat through center of mass diffusion, the contribution of
 which was previously shown
 to be negligible \cite{Keblinski}. However, some recent
experimental high precision studies reported a normal conductivity
in nanoparticle suspensions at very small volume fractions below
1\% \cite{putnam}, questioning the validity of this assumption.
Recent simulations also showed that normal conductivity is
expected for low volume fractions (around $3.3\%$) \cite{Evans} as well as for
volume fractions up to $13\%$ \cite{nanoletter} establishing that the
physical parameter determining thermal properties should be the
particle interfacial thermal resistance.

We also recently established by simulations that Brownian motion
and Brownian velocity field have no effect on the heat transfer
of a single nanoscopic particle with the surrounding fluid \cite{nanoletter}.
Thus the explanation of the increase in conductivity is currently to be
looked for mostly in some collective effects between particles - a field that
has not been studied through microscopic simulation.

In  this work we use non equilibrium molecular dynamics
"experiments" to explore further the transfer of heat in a model
fluid containing nanoparticles.
We make use of the flexibility
allowed by molecular simulations to explore extreme cases in terms
e.g. particle/fluid mass density mismatch. We concentrate on model
systems that are expected to be representative of generic
properties. We explore a large range of parameters and
make a quantitative comparison with effective medium
calculations. By studying the thermal conductivity in a system
with two particles and by precisely controlling their positions
we are able to observe the influence of collective effects consisting
in different particle-particle interactions and displacement with
respect to the temperature gradient.

We start by describing our simulation methodology and presenting
results about the conductivity of a fluid containing a single
nanoparticle in a temperature gradient. We then study the
conductivity of a system containing two particles varying their
positions and the intensity of their interactions.

\section{Simulation Method}
The model fluid used in this study is a simple Lennard-Jones
liquid. The nanoparticles (solid phase) are obtained by a
spherical cut  of a bulk FCC crystal.
 All atoms in our
system interact through Lennard-Jones interactions
\begin{equation}
\label{eq:ljpotcut} U_{lj}(r) =  \bigg\{ \begin{array}{lll}
         4\varepsilon((\sigma/r)^{12} - c(\sigma/r)^{6}), &r\le r_c \\
         0, &r>r_c
       \end{array}
\end{equation}
where $r_c=2.5\sigma$. The coefficient $c$ is equal to 1 for atoms
belonging to the same phase, but can be adjusted to modify the
wetting properties of the liquid on the solid particle
\cite{barratbocquet,barratchiaruttini}. A coefficient of $c=1$ defines
a wetting interaction while the non wetting case is modelled by $c=0.5$.
Within the solid particles, atoms are linked with their neighbors through a
  FENE (Finite extension non-linear elastic) bonding potential:
\begin{equation}
\label{eq:FENE} U_{FENE}(r) =  \frac{k}{2}R_0
\ln(1-(\frac{r}{R_0})^2), \qquad r<R_0
\end{equation}
where $R_0=1.5\sigma$ and $k=30.0 \varepsilon/\sigma^{2}$. This
potential, combined with the Lennard-Jones interaction results in
a narrow distribution of the distance between linked atoms around
$0.97 \sigma$. A solid particle in the fluid was prepared as
follows: starting from a FCC bulk arrangement of atoms at zero
temperature, the atoms within a sphere were linked to their first
neighbors by the FENE bond. Then the system was equilibrated in a
constant NVE ensemble with energy value corresponding to a
temperature $T=1$. A particle contains 555 atoms, surrounded by
the atoms of the liquid. The number density  in  the system is
$\rho = 0.85\sigma^{-3}$. As the simulated particles are not
exactly spherical, but  present some FCC facets, their radius was
estimated from the radius of gyration:
\begin{equation}
\langle R_g^2 \rangle = \frac{1}{N} \sum_1^N (r_i - r_{CM})^2 =
\frac{3}{5}R_p^2
\end{equation}
where $R_g^2$ is the measured radius of gyration of the particle
atoms, and the second equality applies to an ideal sphere. Taking
$\sigma=0.3nm$ this corresponds to a particle radius of order
$R_{part} \sim 1.5nm$. The  solid particles obtained by this
procedure behave very closely as  ideal harmonic solids. Non
linear effects associated with the non linearity of the bonding
potential are absent in a wide temperature range. This assumption
was verified through equilibrium simulations at different
temperatures, monitoring the energy per particle. The observed
relation is linear and indicates a particle heat capacity very
close to $3k_B T N$, as for an harmonic ideal solid, in a
temperature range from $T=1$ to $T=3.5$. We study systems of
wetting ("hydrophilic", $c=1$) and non wetting ("hydrophobic"
$c=0.5$) particles, different values of the density mismatch
between the solid and liquid phase ($m_p/m_l = 1,50,100$ where
$m_l$ is the mass of a fluid atom and $m_p$ is the mass of a
particle atom) and different particle volume fractions.

\subsection{Simulating Heat Flow and Measuring Conductivity}

The most simple and direct method of measuring thermal
conductivities in a simulation is undoubtedly non equilibrium
molecular dynamics. With this set up a temperature gradient is
established through the sample and thermal properties are
calculated from the measurement of energy fluxes. This method
involves locally injecting and evacuating heat in the system. In
our study this is achieved by applying two thermostats with
different temperatures in the two ends of a fluid slab. The
periodicity of the box is maintained in the directions
perpendicular to the temperature gradient and the system is non
periodic in the direction of the heat flux. If one allows the
gradient to change sign in the simulation box a stationary heat
flow can be achieved in a fully periodic box \cite{Evans}. However
with this setup special attention should be payed in the regions
of abrupt change in the temperature gradient, and the simulated
system is less realistic.

\begin{figure}[ht]
\centering
\includegraphics[width=6cm]{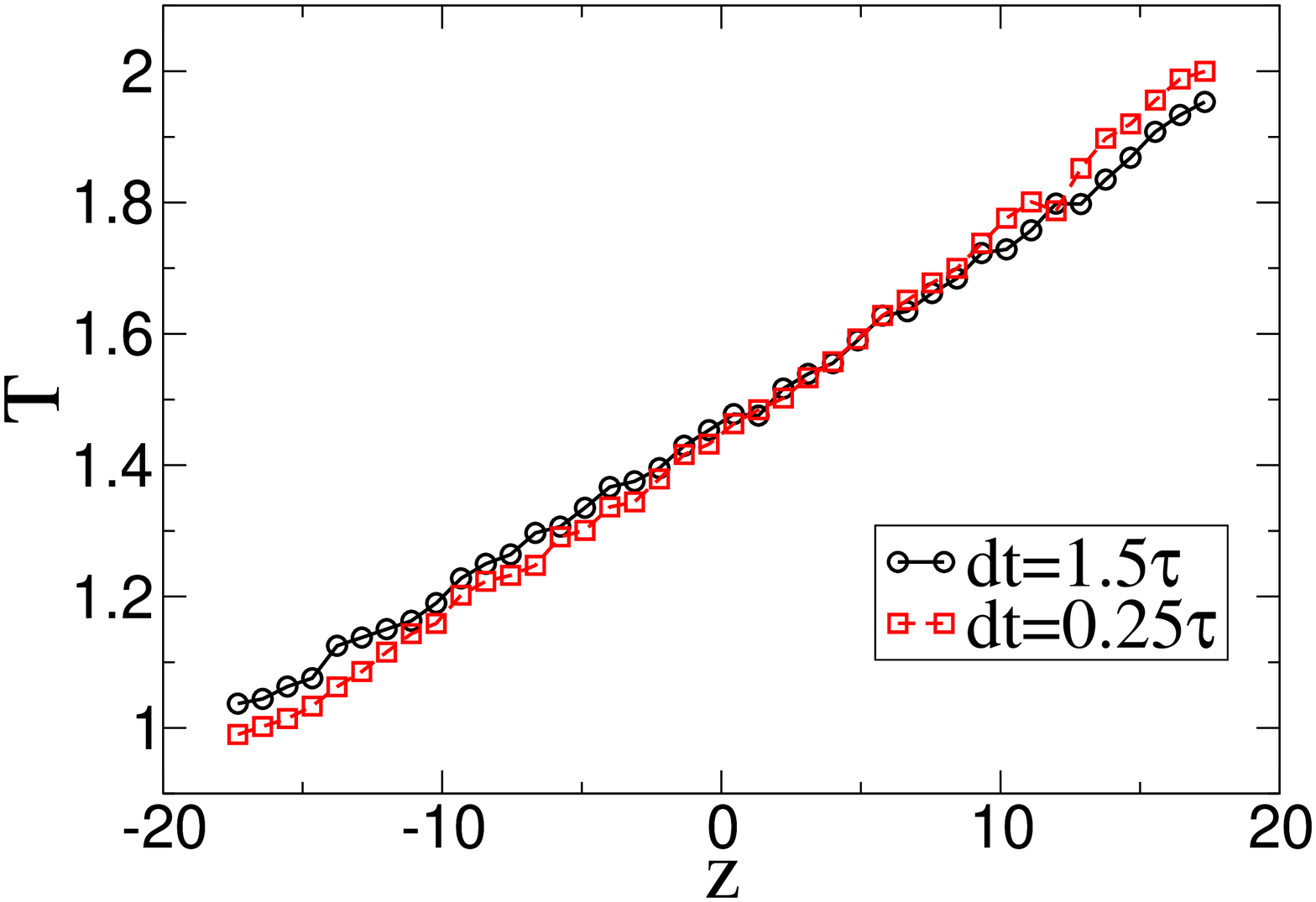}
\caption {Dependence of the temperature profile on the
thermostat time constant for a pure liquid. A smaller rescaling interval
increases the efficiency of the energy transfer between
the thermostat and the liquid inducing a smaller temperature
jump.} \label{fig:tstat-time}
\end{figure}

 The systems we study are periodic in the
$x$ and $y$ directions. In the transverse  $z$ direction, the
liquid is confined by a repulsive potential (ideal flat wall). The
thermostats are applied to a fluid slice in the vicinity of the
walls of width of around 3 atom diameters. The thermostat consists
in a rescaling of the velocities of the particles currently
present in the slice at a given time interval. The temperature
measured locally in slices parallel to the heat flow direction
shows that a linear profile is established in the fluid slab. The
value of the temperature gradient depends on the temperature of
the thermostats and their rescaling time constant (see fig.
\ref{fig:tstat-time}).

A smaller rescaling interval increases slightly the energy
transfer between the thermostat and the fluid thus maintaining the
thermostatted layer at an average temperature closer to the
temperature of the thermostat. It is thus important to have the
same time constant for the thermostat in all the systems in order
to compare the values of the energy flow and detect its variations
in the different setups. We set this value to $dt=1.5\tau$ in the
simulations. Using this heat transfer setup, the local thermal
conductivity is given by:
\begin{equation}
\lambda(z) = \frac{J(z)}{\nabla T(z)}
\end{equation}
where $J(z)$ is the heat flow and $T(z)$ the temperature. This formula
is easily generalized to the whole slab by taking the heat flow at the
thermostats and a mean slope of the temperature profile. The problem
is that such an estimation of the mean gradient is prone to error. The
mean temperature profile for a pure liquid has a well defined slope,
even if the fluctuations and the fitting procedure cause error. The
situation is worse in the presence of nanoparticles. The temperature
profile averaged over liquid and solid atoms has noticeable
fluctuations around the particles as the gradient is different in the
two phases. That is why measuring an effective value of the
conductivity that does not involve fitting and assuming linear
temperature profile is preferable. If the thermostat relaxation
time is constant for all systems, a well defined values are the two thermostats
temperatures. Thus we define the slab conductivity as:
\begin{equation}
\lambda_{eff} = \frac{J}{(T_1 - T_2)/L}
\end{equation}
where $J$ is the mean stationary heat flux measured by the
thermostats, $T_i$ is the temperature of the thermostat $i$ and
$L$ is the length of the simulation box in the $z$ dimension. The
conductivity defined in  this way is  sensitive only  to
variations in the energy flux and is a  precise method for
capturing variations in conductivities between systems.

To avoid any effect of thermophoresis or coupling of the
thermostat to the particle, the particles are constrained to stay
away from the thermostatted regions by tethering weakly their
center of mass to a fixed point by harmonic springs of stiffness
$k=30$. Controlling the particles position also allows us to study
different configurations and explore the influence of the spatial
distribution of the particles on the thermal properties. The
temperatures of the two thermostats were $T_1 = 2$ and $T_2 = 1$.
In order to compare the conductivity results for the different
systems they were first equilibrated to the same pressure at a
temperature $T=1.5$, then a non equilibrium run was performed for
about 1500-2000 $\tau_{LJ}$ to make sure the pressure stays the
same for the systems of different nature and finally a production
run of about 15000 $\tau_{LJ}$ during which thermostats energy,
particle positions and temperature profiles are monitored. In
nanoscale systems it has been observed that interfacial effects
are very important \cite{nanoletter}. In order to compare
quantitatively the conductivity variations to the predictions of
effective medium calculations it is necessary to know the value of
the interfacial (Kapitza) thermal resistance. The values used in
our study were determined for different wetting and particle
masses by a transient adsorption simulation as explained in ref.
\cite{nanoletter}. In real units, a value $R_K=1$ corresponds
typically to an interfacial \textit{conductance} $G=1/R_K$, of the
order of $100\mathrm{ MW/K m}^2$ \footnote{The conversion to
physical units is made by taking a Lennard-Jones time unit
$\tau_{LJ}=10^{-12}s$, and a length unit $\sigma=0.3nm$. The unit
for $G$ is $energy/temperature/(length)^2/time$. As the
energy/temperature ratio is given by the Boltzmann constant $k_B$,
we end up with a unit for $G$ equal to $k_B /\sigma^2 /\tau_{LJ}
\simeq =10^8 W/m^2/s$}. We study systems with particle volume
fraction of $13\%$ or $12\%$. The volume fraction is defined as
the volume of the particle divided by the volume of the fluid
outside the thermostats.

\section{Results and Discussion}

\subsection{Single Particle in a Heat Flux}
\begin{figure}[ht]
\centering
\includegraphics[width=6cm]{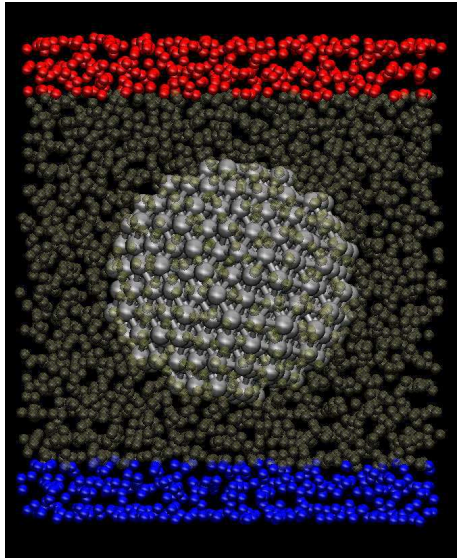}
\caption[Snapshot of heat transfer setup] {Snapshot of the
system used to evaluate the thermal conductivity with a
particle of 13\% volume fraction.} \label{fig:snap1}
\end{figure}

First we investigate the effect of the presence of a single
nanoparticle on the thermal conductivity of the fluid. In this
system the  center of mass of the particle is held at equal
distance from the two thermostats (fig. \ref{fig:snap1}).

\begin{figure}[ht]
\centering
\includegraphics[width=7cm]{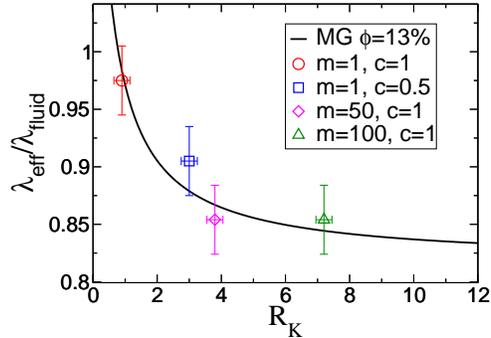}
\caption[Comparison of simulation results and MG equation]
{Comparison between the ratio of the effective conductivity to the
conductivity of the pure liquid of the simulated systems and the
values obtained from the Maxwell Garnett equation.}
\label{fig:mg}
\end{figure}

We performed simulations for a
small volume fraction ($\Phi \sim 2\%$), where we were not able to
detect any change in thermal conductivity compared to the bulk
fluid.  At a higher volume fraction ($\Phi \sim 13\%$), on the
other hand, we observe a clear {\it decrease} in the heat
conductivity associated with the presence of the nanoparticle
(fig. \ref{fig:mg}).
The Kapitza resistance $R_K$ for the considered
particles ranges from 1 to 7 so that the associated characteristic
Kapitza length is in all cases of the order of the particles diameter.
This means that the decrease must be interpreted in terms
of interfacial effects. To quantify these effects, we use the
 Maxwell-Garnett approximation for
spherical particles, modified to  account account for the Kapitza
resistance at the boundary between the two media. The resulting
expression for the effective conductivity \cite{MG} (see appendix)
 is
\begin{equation}
\frac{\lambda_{eff}}{\lambda_l}= \frac{\big(
\frac{\lambda_p}{\lambda_l}(1+2\alpha)+2 \big) + 2\Phi\big(
\frac{\lambda_p}{\lambda_l}(1-\alpha)-1 \big)} {\big(
\frac{\lambda_p}{\lambda_l}(1+2\alpha)+2 \big) - \Phi\big(
\frac{\lambda_p}{\lambda_l}(1-\alpha)-1 \big)}
\end{equation}
where $\lambda_l$ and $\lambda_p$ are the liquid and particle
conductivities, $\Phi$ is the particle volume fraction and $\alpha
= \frac{R_K \lambda_l}{R_p}$ is the ratio between the Kapitza
length (equivalent thermal thickness of the interface) and the
particle radius. This model predicts an increase in the effective
conductivity  for $\alpha
> 1$  and a
decrease for $\alpha < 1$, regardless of the value of the
conductivity of the particles or of the  volume fraction. The
prediction depends very weakly on the ratio $\lambda_p/\lambda_l$,
less than 1\% for $10<\lambda_p/\lambda_l<100$.
 The minimum value   of
$\lambda_{eff}/\lambda_l$, obtained when $\alpha \to \infty$, is
$\frac{1-\Phi}{1+\Phi/2}$ while the maximum possible enhancement
(for $\lambda_p \to \infty$ and $R_K \to 0$) is
$\frac{1+2\Phi}{1-\Phi}$. The Kapitza resistance can be modified
by tuning either the liquid solid interaction coefficient $c$, or
the mass density of the solid, or a combination of these two
parameters. Figure \ref{fig:mg} illustrates the variation of the
measured effective conductivity for several values of the Kapitza
resistance, determined independently for various values of these
parameters. It is seen that the observed variation (decrease in
our case) in the effective conductivity is very well described by
the Maxwell-Garnett expression. This expression also allows us to
understand why the heat conductivity does not vary in a
perceptible manner for small volume fractions ($\sim 2\%$), for
which the predicted change would be less than $2\%$, within our
statistical accuracy.

\subsection{Study of Collective Effects}
\begin{figure}[ht]
\centering
\includegraphics[width=7cm]{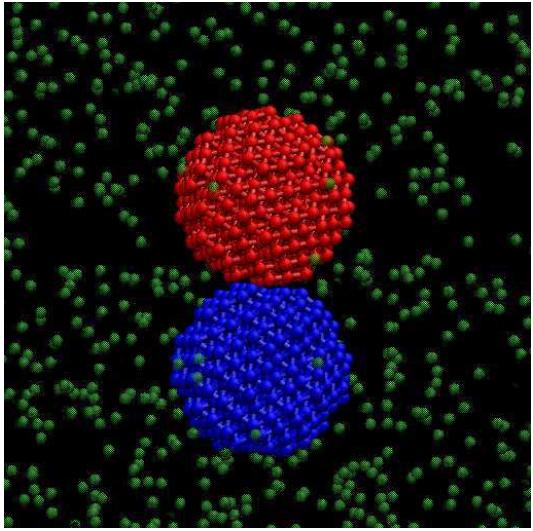}
\caption {Snapshot of the
system containing two particles aligned with the temperature
gradient having strong particle-particle interaction.}
\label{fig:snap2}
\end{figure}

Next we study a system of volume fraction $12\%$ containing two
nanoparticles in various configurations, in order to investigate
 the influence of collective effects. In order to model
the influence of microscopic particle clustering we study two
particles tethered by soft springs and forced  to stay in
``contact'', directly interacting with each other (fig.
\ref{fig:snap2}. We modify the position of the  centers of mass of
the particles with respect to the temperature gradient - the
particles are either aligned parallel or perpendicular to the
gradient. In these two situations the solid phase has an aspect
ratio of either $a=2$ (parallel) or $a=1/2$ (perpendicular). We
also varied the Lennard-Jones interaction intensity between the
two particles ($\varepsilon_{pp}$) in order to increase the rate
of particle-particle energy transfer. The particles-liquid thermal
resistance is also modified by choosing different wetting
properties or different ratios of the masses of the atoms
belonging to the two phases. Finally, to explore a broader range
of configurations, we also modified the distance between the
centers of mass of the particles. In most configurations the
particles are directly interacting with each other and in two
setups they are separated along  the $z$ direction by a layer of
atoms from the liquid phase, with a thickness of the order of a
particle diameter . An overview of the systems investigated is
shown in table \ref{table:systems}.
\begin{table}[ht]
\centering
\begin{tabular}{|l||c|c|c|c|c|c|}
\hline
System name & Aspect ratio $a$ & $\varepsilon_{pp}$ & $m_p/m_l$ & $c$
& $R_K$ & Particle-particle $\Delta r$ \\ \hline
$a2\varepsilon 1m1c1$ & 2.0 & 1.0 & 1 & 1 & 0.8 & $2R_p$ \\
$a2\varepsilon 10m1c1$ & 2.0 & 10.0 & 1 & 1 & 0.8 & $2R_p$ \\
$a2\varepsilon 1m50c1$ & 2.0 & 1.0 & 50 & 1 & 4.0 & $2R_p$ \\
$a2\varepsilon 10m50c1$ & 2.0 & 10.0 & 50 & 1 & 4.0 & $2R_p$ \\
$a2\varepsilon 10m1c0.5$ & 2.0 & 1.0 & 1 & 0.5 & 3.2 & $2R_p$ \\
$a0.5\varepsilon 1m1c1$ & 0.5 & 1.0 & 1 & 1 & 0.8 & $2R_p$ \\
$a0.5\varepsilon 10m1c1$ & 0.5 & 10.0 & 1 & 1 & 0.8 & $2R_p$ \\
$a0.5\varepsilon 1m50c1$ & 0.5 & 1.0 & 50 & 1 & 4.0 & $2R_p$ \\
$a0.5\varepsilon 10m50c1$ & 0.5 & 10.0 & 50 & 1 & 4.0 & $2R_p$ \\
$\varepsilon 1m1c1$ & $\times$ & 1.0 & 1 & 1 & 0.8 & $4R_p$ \\
$\varepsilon 1m50c1$ & $\times$ & 1.0 & 50 & 1 & 4.0 & $4R_p$ \\
\hline
\end{tabular}
\label{table:systems} \caption{Description  of the  systems
containing two particles . The
  last column indicates the distance between the centers of mass of
  the two particles.}
\end{table}

First we study systems with  aspect ratio $a=2$ and strong
particle-particle interaction, as a function of the interfacial
resistance with th efluid(fig. \ref{fig:snap2}). The strong
interaction is essentially equivalent to a chemical bonding
between the particles.  Te two particles move as a single rigid
body and the heat transfer between them is considerably enhanced.
We can therefore reasonably   compare the obtained conductivities
with the values  an effective medium calculation would predict for
an ellipsoid with the same aspect ratio.
A calculation including interfacial thermal resistance for
ellipsoidal particles of the same size, aligned with the thermal
gradient, relevant for this situation can be derived from a
general expression in ref. \cite{MG}. The ratio of the effective
conductivity to the conductivity of the pure liquid is given by:
\begin{equation}
\frac{\lambda_{eff}}{\lambda_l}= \frac{1+\Phi \beta (1-L_{zz})}
{1-\Phi \beta L_{zz}}
\label{eqn:asp}
\end{equation}
where $\Phi$ is the volume fraction, $L_{zz}$ is given by
\begin{equation}
L_{zz}=1-2\left( \frac{a^2}{2(a^2-1)} -
  \frac{a^2}{2(a^2-1)^{3/2}}\cosh^{-1}a \right)
\end{equation}
with aspect ratio $a>1$. The parameter $\beta$ is given by
\begin{equation}
\beta=\frac{\lambda^c - \lambda_l}{\lambda_l + L_{zz}(\lambda^c - \lambda_l)}
\end{equation}
where
\begin{equation}
\lambda^c = \frac{\lambda_p}{1+\frac{\lambda_p}{\lambda_l}
(2+1/a)\frac{R_K\lambda_l}{R_p}L_{zz}}
\end{equation}
In the above $\lambda_p$ and $\lambda_l$ are the conductivities of the
particles and the liquid, $R_p$ is the particles radius and $R_K$ -
the Kapitza resistance of the particle-liquid interface. The relations
hold for an aspect ratio $a>1$.

\begin{figure}[ht]
\centering
\includegraphics[width=7cm]{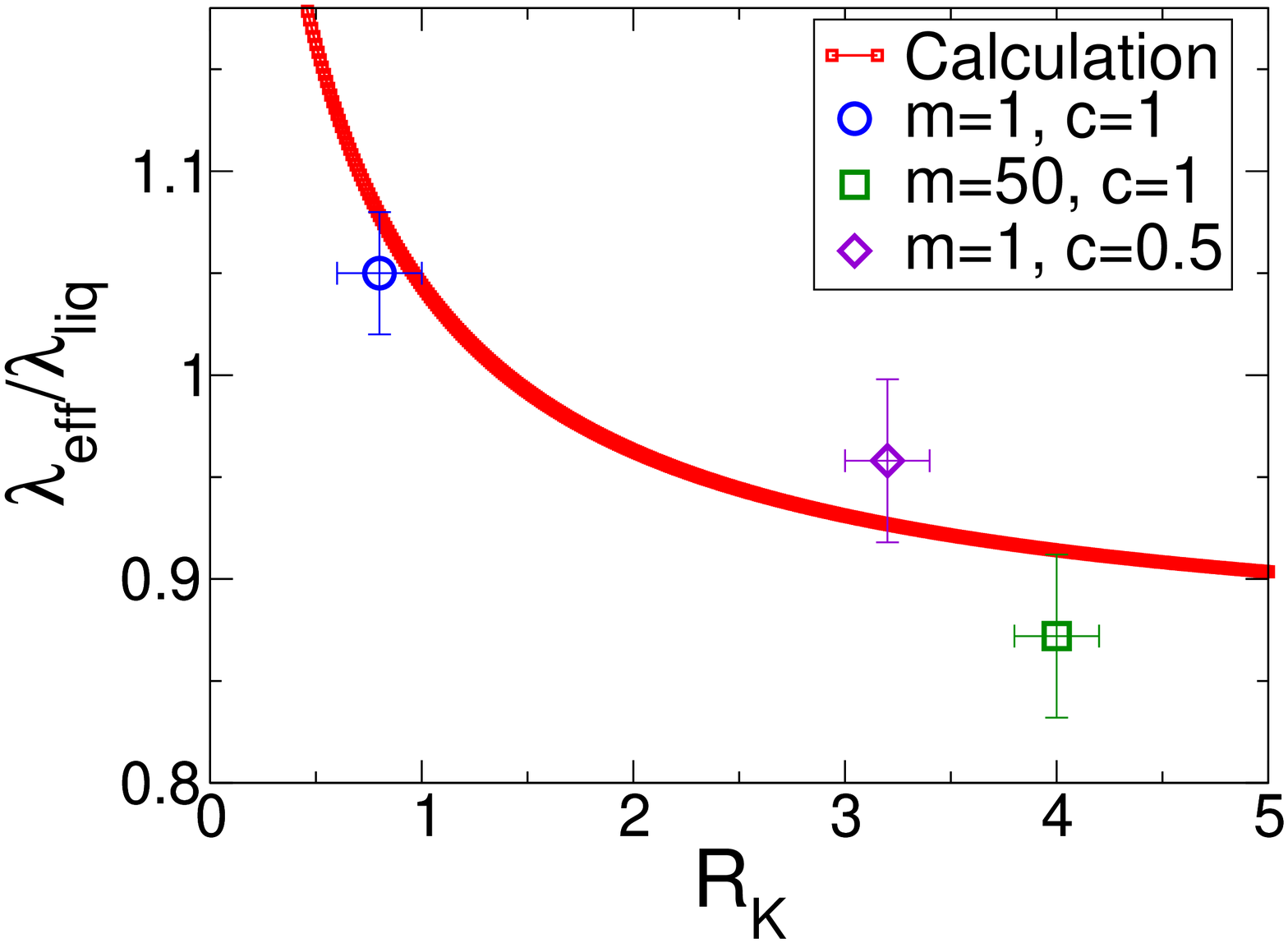}
\caption{Comparison of the measured variation in conductivity
in the $\varepsilon =10$, $a=2$ systems to the prediction
of the effective medium calculation with aspect
ratio 2 and volume fraction 0.12 (equation \ref{eqn:asp}).}
\label{fig:mg-asp}
\end{figure}

As can be seen in figure \ref{fig:mg-asp} the measured
conductivities are in reasonable agreement with the calculation.
For the smallest interfacial resistance there is an enhancement of
the composite system conductivity  by $5$ to $10\%$. For larger
values of  $R_K$ the small radius of the particles  results in a
decrease in the conductivity, due to interfacial effects.

\begin{figure}[ht]
\centering
\includegraphics[width=9cm]{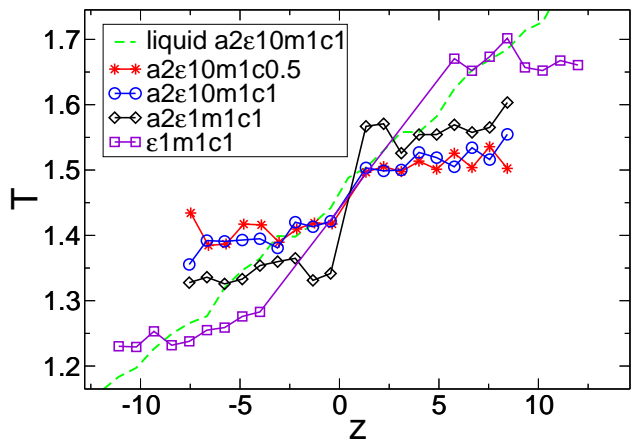}
\caption{Temperature profiles in the particles in the $m=1$ systems
  in stationary heat flow. The temperature profile within the
  liquid  is also shown for one of the systems. For the other systems
  the liquid temperature profile is very similar. Stronger interaction
  decreases the temperature difference between the particles by
  approximately 50\% compared to particles
  separated by the same distance with neutral interaction.
}
\label{fig:temp}
\end{figure}

Next, we study the thermal behavior of the systems with particles
aligned with the thermal gradient but much weaker interactions.
The  particle-particle interaction  is taken to be neutral,
$\varepsilon_{pp} = 1$ (configurations $a2\varepsilon 1m50c1$,
$a2\varepsilon 1m1c1$ in table \ref{table:systems}). For this
weaker interaction, we find that the   conductivity is typically
$4\%$ below the one obtained with the strong attractive
interaction. This difference can be understood from the
temperature profiles shown in figure \ref{fig:temp}. The weaker
interaction results in a higher resistivity in the "neck" region,
at the boundary  between the two particles.
As the strongly attractive interaction decreases the
particle-particle thermal resistance, the particles temperatures
are closer for $\varepsilon_{pp}=10$. As the particles are much
more conductive than the liquid phase their temperature varies
very little within their dimension. We observed that the
temperature difference between the particles decreases by a factor
of two when the interaction intensity is increased by a factor of
ten: $\Delta T = 0.22$ for $\varepsilon_{pp}=1$ and $\Delta T =
0.12$ for $\varepsilon_{pp}=10$ in the systems of $m=1$. In the
$m=50$ case $\Delta T (\varepsilon_{pp}=1)= 0.44$ and $\Delta T
(\varepsilon_{pp}=10)= 0.2$. Because of the layers of fluid (of
thickness around $5\sigma$) remaining in both cases between the
particles and the thermostats the global conductivity of the slab
is still dominated by interfacial effects and its increase is
small.

When the particle-particle distance is increased, so that the
particles do not interact directly with each other, the
conductivity of the sample stays nearly identical (difference
$\sim 1\%$ in favor for the system where the particles are closer)
to the case where they are in contact and with neutral
interaction. Knowing that the contact area is very small when the
particles are close and also given that the interaction is not
strong, being equal to the interaction with the liquid atoms
$\varepsilon_{pp}= \varepsilon = 1$, the inter particle distance
in this case does not play an important role in the value of the
conductivity. As can be seen in fig. \ref{fig:temp}, whenever
$\varepsilon_{pp}= \varepsilon = 1$ the average temperature of the
particle is close to the average temperature of the liquid at the
$z$ coordinate of its center of mass, the particle thermalizes
with the surrounding fluid. In contrast, when
$\varepsilon_{pp}=10$ the inter particle heat flux becomes
important and the two particles behave more like a single body.


We now turn to the conductivity of the systems where the particles are
in the plane situated in the middle of the box and orthogonal to the
temperature gradient. The heat fluxes measured for such  systems of
aspect ratio 1/2, with
strong and neutral particle-particle interaction, were identical within
our statistical accuracy (difference less than one percent). The
conductivities in this case are slightly lower, with about $4.5\%$
($2\%$ for $m=1$ and $7\%$ for $m=50$) than in the case where the
particles are aligned with the temperature gradient and has neutral
interactions ($\varepsilon_{pp}=1$). The difference is further
increased to $\sim 8\%$ ($7\%$ for $m=1$ and $10\%$ for $m=50$) when
the $a=0.5$ systems are compared with the $a=2$ and
$\varepsilon_{pp}=10$ systems.

\begin{figure}[ht]
\centering
\includegraphics[width=7cm]{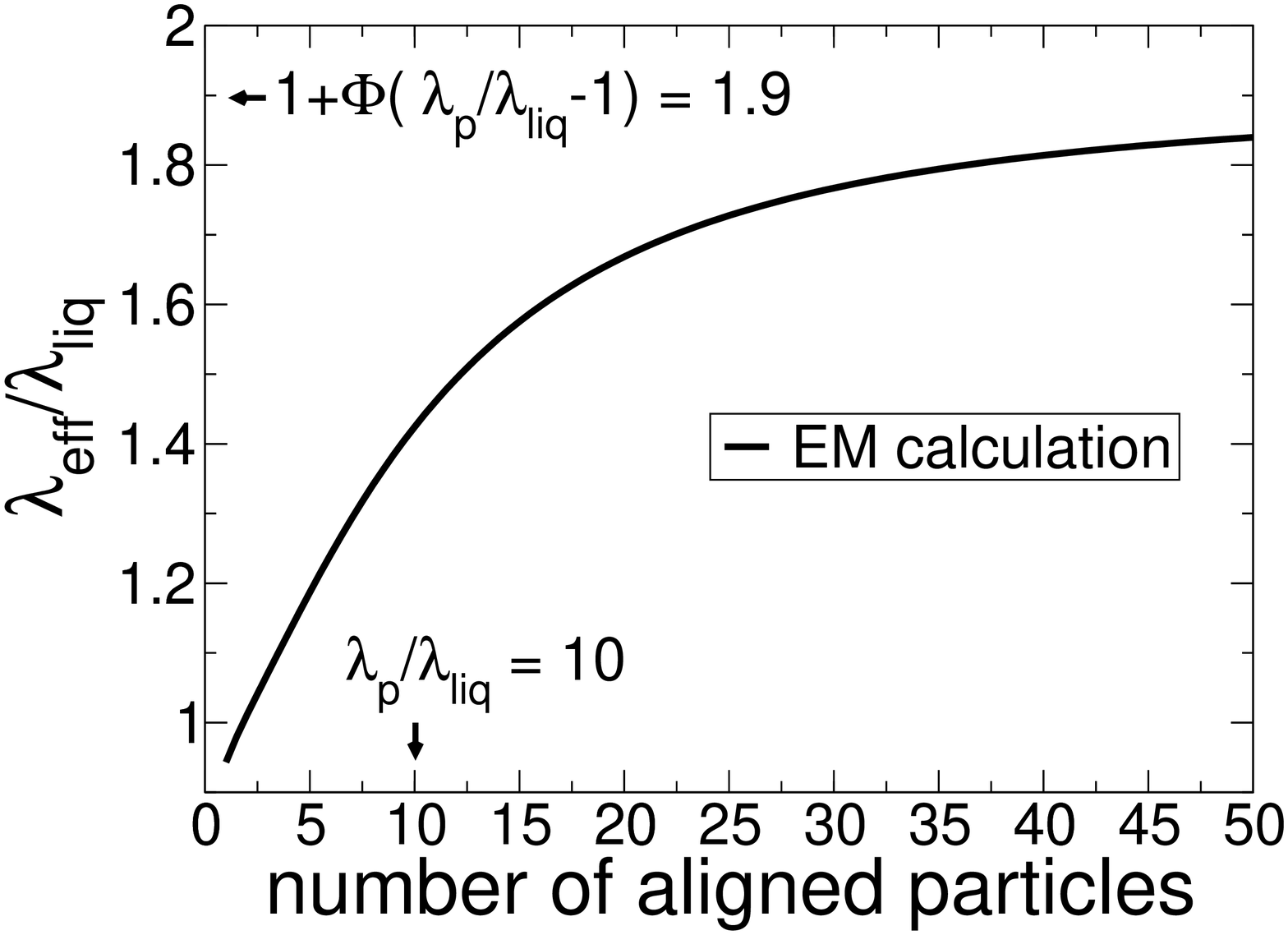}
\caption{Prediction of the effective medium calculation (equation
  \ref{eqn:asp}) for the ratio of the system conductivity to the
  conductivity of the liquid for an interfacial resistance $R_K=1$,
  a volume fraction $\Phi=0.1$, and a ratio of $10$ between the thermal conductivities
  of the solid and of the liquid.
  .}
\label{fig:asp-th}
\end{figure}

In summary, we showed that the mutual positions of the particles in
suspension in the fluid has an influence on the thermal conductivity of
the system. If the particles interact so that clustering occurs in the
suspension, the global conductivity of the nanofluid can increase to
values higher than the one of  a pure fluid, even if a well dispersed
suspension at the same volume fraction has a conductivity lower than
the pure system. The alignment with the temperature gradient enhances
the conductivity and its effect on a microscopic level can be
predicted by effective medium calculation taking into account the
aspect ratio of the particle cluster. According to the effective
medium prediction the ratio of the slab conductivity to the
conductivity of the pure liquid grows essentially linearly with the number of
aligned particles (or the aspect ratio) (see fig. \ref{fig:asp-th})
when this number is smaller than the ratio of the conductivities of
the two phases ($\lambda_p/\lambda_l$). Hence even a moderate clustering (e.g.
in strings of 3 to 4 particles) could be sufficient to interpret increases in the
thermal conductivity compared to standard effective medium predictions.

The enhancement tends to
$(1-\Phi) + \Phi\lambda_p/\lambda_l$ when the aspect ratio tends to
infinity. Thus, following this calculation that seems to give reasonable
results at least on the scale of two particles and given that the ratio
$\lambda_p/\lambda_l$ is usually high in real nanofluids, a large
enhancement can be expected if aggregates
 of particles are formed in the suspension. Our calculation concerns string like aggregates,
 but it is likely that similar effects would be observed with other types of aggregates,
 e.g. fractal ones with  a low enough dimension.

\section{Conclusion}
We have explored some  aspects of the thermal properties
of "nanofluids", at the level of model system, individual solid
particles and collective effects involving two particles on a
microscopic scale.

By varying interaction parameters or mass
density, we are able to vary the interfacial resistance between
the particle and the fluid in a large range. This allowed us to
estimate, over a large range of parameters, the effective heat
conductivity of a model nanofluid in which the particles would be
perfectly dispersed. The results for this setup can be simply explained in terms
of the classical Maxwell-Garnett model, provided  the interfacial
resistance is taken into account. The essential parameter that
influences the effective conductivity turns out to be the ratio
between the Kapitza length and the particle radius, and for very
small particles a decrease in conductivity compared to bulk fluids
is found.

We also examined the effect of particle clustering and alignment with
respect to the temperature gradient. We found that alignment improves
the conductivity of the nanofluid in accordance with calculation with
effective medium approach. Increased inter particle interaction further
enhances conductivity. The results for small clusters of particles can again be
described by effective medium theory, taking into account anisotropy and interface effects.
Clustering of particles into string like objects is therefore suggested as a possible
mechanism for obtaining larger conductivities, compared to the case of well dispersed suspensions.

In order to draw more precise conclusions concerning the conductivity
dependence on the complex physics of collective effects
larger systems containing more complex aggregates
should be examined. The present study provides guidelines and
outlines general tendencies that are to be expected in a larger and
more complex system.

\section{Appendix}
\begin{figure}[ht]
\centering
\includegraphics[width=7cm]{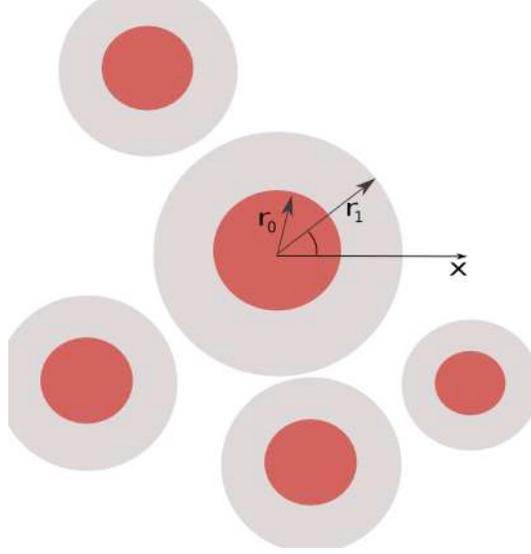}
\caption{Schematic presentation of the matrix with dispersed inclusions}
\label{fig:mg-sch}
\end{figure}

We present  here a brief  derivation of the Maxwell Garnett equation
for the effective thermal conductivity in a two phase media (matrix
with spherical inclusions) taking into account the interfacial thermal
resistance. We consider a macroscopically homogeneous material with
thermal conductivity $\lambda_{eff}$ in a temperature gradient
following some axis: $T_{eff}(r)=- \mathbf{g \cdot r}$. We concentrate on a spherical
inclusion of radius $r_0$ and conductivity $\lambda_1$ surrounded by a
spherical shell of host material of thickness $r_1$, matrix conductivity
$\lambda_2$ and we assume that the inclusion does not change the
temperature field for $r>r_1$. The two radii define the volume fraction of
the inclusion, $\Phi = r_0^3/r_1^3$ (fig. \ref{fig:mg-sch}).
The effective conductivity can be determined by solving the steady
state diffusion equation for the temperature:
\begin{equation}
\Delta T(r,\theta)=0
\label{eqn:diff}
\end{equation}
where $\theta$ is the angle between $\mathbf{r}$ and the external
gradient $\mathbf{g}$. The solutions of equation \ref{eqn:diff} in the
three different regions are given by:
\begin{eqnarray}
T_1(r,\theta ) &=& Ar\cos \theta ,\; 0<r\leq r_0 \\
T_2(r,\theta ) &=& ( Br+ E/r^2)\cos \theta ,\; r_0<r\leq r_1\\
T_{eff}(r,\theta ) &=& -gr\cos \theta ,\; r_1<r
\end{eqnarray}
The unknown constants in the equations above are to be determined from
the appropriate boundary conditions. These are obtained by expressing
the continuity of the heat flow at the domain boundaries and the
value of the temperature field:
\begin{eqnarray}
T_1(r_0,\theta ) - T_2(r_0,\theta ) &=& -\lambda_1 \frac{\partial
  T_1}{\partial r}(r_0) R_K\\
\label{eqn:bc1}
\lambda_1 \frac{\partial  T_1}{\partial r}(r_0) &=& \lambda_2 \frac{\partial
  T_2}{\partial r}(r_0) \\
T_2(r_1,\theta ) &=& T_{eff}(r_1,\theta) \\
\lambda_2 \frac{\partial  T_2}{\partial r}(r_1) &=& \lambda_{eff} \frac{\partial
  T_{eff}}{\partial r}(r_1)
\label{eqn:bc2}
\end{eqnarray}
where $R_K$ is the interfacial thermal resistance responsible for a
temperature jump at the matrix-inclusion interface. Substituting the
solutions for the temperature fields in equations
\ref{eqn:bc1}-\ref{eqn:bc2}, we end up with the following relations:
\begin{eqnarray}
Ar_0^3 - Br_0^3 - E + \lambda_1R_KAr_0^2 &=& 0 \\
\label{eqn:bc-sol1}
\lambda_1Ar_0^3-\lambda_2Br_0^3+2\lambda_2E &=& 0 \\
Br_1^3 + E+gr_1^3 &=& 0 \\
\lambda_2Br_1^3 - 2\lambda_2E + \lambda_{eff}gr_1^3 &=& 0
\label{eqn:bc-sol2}
\end{eqnarray}
Using this set of equations, one can obtain the Maxwell Garnett
equation for the effective conductivity with interfacial thermal
resistance:
\begin{equation}
\frac{\lambda_{eff}}{\lambda_2}= \frac{\big(
\frac{\lambda_1}{\lambda_2}(1+2\alpha)+2 \big) + 2\Phi\big(
\frac{\lambda_1}{\lambda_2}(1-\alpha)-1 \big)} {\big(
\frac{\lambda_1}{\lambda_2}(1+2\alpha)+2 \big) - \Phi\big(
\frac{\lambda_1}{\lambda_2}(1-\alpha)-1 \big)}
\end{equation}
where $\alpha = R_K\lambda_2/r_0$.

\bibliographystyle{unsrt}

\end{document}